\documentclass[prl,preprintnumbers,twocolumn,aps,superscriptaddress]{revtex4}
\usepackage{graphicx,}
\usepackage{graphicx,dcolumn,bm,epsfig,amsmath,amssymb,}
\usepackage{booktabs}
\usepackage{multirow}
\usepackage{float}

\begin{document}

\title {Re-visiting gravitational wave events via pulsars}

\author{Minati Biswal}
\email{minati.b@iopb.res.in}              
\affiliation{Institute of Physics, Bhubaneswar 751005, India}

\author{Shreyansh S. Dave}
\email{shreyanshsd@imsc.res.in}
\affiliation{The Institute of Mathematical Sciences, Chennai 600113, India}

\author{Ajit M. Srivastava}
\email{ajit@iopb.res.in}
\affiliation{Institute of Physics, Bhubaneswar 751005, India}
\affiliation{Homi Bhabha National Institute, Training School Complex,
Anushakti Nagar, Mumbai 400085, India}

\begin{abstract}
By now many gravitational wave (GW) signals have been detected by LIGO
and Virgo, with the waves reaching earth directly 
from their respective sources. These waves will also travel to different
pulsars and will cause (tiny) transient deformations in the pulsar shape. 
Some of us have recently shown that the resultant transient change in 
the pulsar moment of inertia may leave an observable imprint on the pulsar 
signals as detected on earth, especially at resonance. The pulsars may 
thus act as remotely stationed Weber gravitational wave detectors. This
allows us to revisit the past GW events via pulsars.
We give here a list of specific pulsars whose future 
signals will carry the imprints of past GW events, to be specific
we constrain it within 50 years. Some interesting cases are, supernova 
SN1987A with earliest perturbed signals from pulsars J0709-5923  and B0559-57 
expected to reach earth in 2023 and 2024 respectively, Crab supernova,
with perturbed signal arrival date from pulsar J1856-3754 in 2057, 
and GW170814 event with its imprints on the signals on the pulsar J0437-4715 
reaching earth between 2035-2043. Even the earliest recorded supernova SN185 
event may become observable again via pulsar J0900-3144 with the perturbed
pulsar signal reaching us sometime between 2033-2066. Importantly, even though
the strength of the signal will depend on the interior properties of the
pulsar, the expected dates of signal arrival are completely model independent,
depending only on the locations of the source and the relevant pulsar.
\end{abstract}
\maketitle

\section{Introduction}
\noindent 

Detection of gravitational waves (GW) has opened a new chapter in 
observational astronomy. A number of GW sources have been detected by
LIGO and Virgo \cite{gwtc} ranging from coalescing black holes (BH)
to binary neutron star (BNS) mergers. A BNS merger allows much better 
localization of the source due to electromagnetic radiations accompanying
the GW signal. The same is not true for BH mergers where GW source
localization is very hard with very few GW detectors. Recently some of us 
have proposed \cite{pulsargw} that sensitive monitoring of pulsars 
can be used to detect the arrival of GW signals on those pulsars. 
This possibility arises due to deformations caused by the gravitational 
wave (GW) passing through a pulsar leading to a transient variation in 
its moment of inertia. This affects the extremely accurately measured spin 
rate of the pulsar as well as its pulse profile (due to
induced wobbling depending on the source direction). The effect will be
most pronounced at resonance. We thus proposed that the pulsars, in 
this sense, act as remotely stationed Weber detectors of gravitational 
waves whose signal can be monitored on earth \cite{pulsargw}.  

 This allows for a new possibility of detecting signals of
GW events (or any other  astrophysical event affecting pulsar moment of 
inertia such as phase transitions occurring inside a pulsar as discussed
in \cite{nstarpt}). Further, this allows us the possibility of re-visiting
past GW events, e.g. those which have already been detected by GW detectors.  
This can give further information about the GW source, and also new information
about the relevant pulsar interior (in the manner its pulses carry the
imprints of the GW signal). This also will allow for better triangulation
of the original source location which is of crucial importance for
black hole mergers which only emit gravitational waves.
Particularly important are the cases of those events whose GW signal 
detection has been missed. For example, many supernova events have been
identified within our galaxy from astronomical records around the world.
Any GW emission from these events should have relatively strong effects 
on the galactic pulsars. (Especially for type-II supernova \cite{sngw} 
with estimated GW strain amplitude reaching as high as $10^{-20}$ at a 
distance of 10 kpc from the source \cite{sngw}, though even for
Type IA supernova the GW signal may be strong \cite{gwsn1a}.) Observing these
pulsars in future (or, from the past recorded pulse data) has the
possibility of identifying the signals of these supernova events
hidden in the pulsar signal. 
 In this paper we have analyzed specific GW events which have been
detected so far \cite{gwtc}, as well as specific supernova events which
are available from astronomical records \cite{sndata}. We then consider 
the locations of a list of pulsars (a total of 2659 pulsars) from 
ref. \cite{pulsardata} and calculate the arrival time of perturbed signals 
on earth. We limit listing of  results here to those pulsars whose 
signals will reach earth within 50 years. We mention some 
interesting cases for the earliest signal arrival dates. For supernova 
SN1987A \cite{SN87}, we find that perturbed signals from pulsars 
J0709-5923  and B0559-57 are expected to reach earth on 5/9/2023 and 
7/ 8/2024 respectively, with many more pulsar signals expected in subsequent 
years.  We give the date of earliest signal arrival in days and months also
even though the error estimates (from errors in source and pulsar
location) are in years. This is only meant to reflect the exactly known 
date of the source, and to illustrate the possibility that with improvements 
in various observations (source and pulsar locations), in principle, a very 
precise date of signal arrival on earth could be predicted. We will
give error estimates, wherever available, in the table below. For Crab 
supernova \cite{crab}, we find the signal arrival date from pulsar 
J1856-3754 in March 2057 with error of one year (from the error in Crab 
source distance).  Among the merger events detected by LIGO/Virgo, 
GW170814 event will have its imprints on the signals on the pulsar 
J0437-4715 reaching earth between 2035-2043.
 
 We first recall basic results from ref. \cite{pulsargw}. Deformations
in the shape of neutron star (NS) result from the changes in the Riemann 
curvature tensor $R_{\mu\nu\lambda\rho}$ as the metric undergoes periodic
variations due to gravitational wave passing through the NS. This induces a
change in its quadrupole moment $Q_{ij}$ which can be written in the following 
form \cite{hinderer2008}:
\begin{equation}
Q_{ij} = - \lambda_d E_{ij}~,
\label{eq:qij}
\end{equation}
Here $E_{ij}$ is the external tidal field. $\lambda_d$ is the tidal 
deformability which can be expressed in terms of the radius $R$ of the
neutron star and the second love number $k_2$,
\begin{equation}
\lambda_d = \frac{2}{3} k_2 \frac{R^5}{G}.  
\label{eq:lambda}
\end{equation}
 The value of $k_2$ is constrained to be in the range $k_2 
\simeq 0.05 - 0.15$ from the observations of the recent BNS merger 
\cite{abbott2017}.
By writing $E_{ij} = R_{i0j0}$ in terms of the GW strain amplitude $h$ for 
a specific polarization and using transverse traceless (TT) gauge,
one gets \cite{carrollbook}:
\begin{equation}
E_{xx} = - E_{yy} = \frac{2\pi^2 h c^2}{\lambda^2},
\label{eq:ett}
\end{equation}
Here $\lambda$ is the wavelength of GW, and we use $h$ to denote the GW 
strain amplitude $h_+$ for the $+$ polarization. By taking an initial 
spherical shape and considering ellipsoidal deformation, Eq.(1) gives the
resulting change in the quadrupole moment tensor. Using that we calculate
the change in the moment of inertia tensor for a neutron star of mass $M$
as \cite{pulsargw}

\begin{equation}
\frac{\Delta I_{xx}}{I} = - \frac{\Delta I_{yy}}{I}
\simeq \frac{k_2}{3} \frac{R^3 c^2}{GM \lambda^2} 20h
\label{eq:deltami}
\end{equation}
We take sample values with 
$M = 1.0 M_\odot$ and $R = 10$ km, and use $k_2$ = 0.1. To get high
sensitivity $\lambda$ needs to be small, at the same time validity 
of Eqn.(1) requires $\lambda$ to be much larger than NS radius $R$.
For the astrophysical source of GW, we consider binary NS merger (such 
as the one observed by LIGO and Virgo) with the highest value of 
GW frequency being about 1 kHz (thus retaining the validity of Eqn.(1)). 
We then get,
\begin{equation}
\frac{\Delta I_{xx}}{I} \simeq 10^{-2} h~.
\label{eq:sampledeltami}
\end{equation}

The BNS merger detected by LIGO-Virgo had peak strength of the signal 
$h \simeq 10^{-19}$ \cite{abbott2017} with the source distance estimated 
to be about 130 million light years. Considering the prototype pulsar 
detector to be at a distance of, say,  a thousand light years from the
GW source, resulting value of $h$ at the pulsar-detector location will get 
enhanced to $h \simeq 10^{-14}$.  (With most pulsars in our galaxy located 
inside dense globular clusters, such a situation is not very unlikely. As we 
emphasized in \cite{pulsargw},
this should provide motivation for detailed and precision studies 
of pulsars very far away, especially extra-galactic pulsars.

Fractional changes in the spin rate $\nu$ of the 
pulsar will be the same as the fractional change in the MI. Thus, we 
estimate changes in the spin rate of the pulsar for this case to be
\begin{equation}
\frac{\Delta \nu}{\nu} = \frac{\Delta I}{I} \simeq 10^{-16}~.
\end{equation}
Here, we have written $\Delta I$ for the change in the relevant component
of MI.  Pulse timings of many millisecond pulsars  have been measured to 
an accuracy better than $10^{-15}$ seconds. Thus spin rate changes 
of order given above are close to the detectable limits by precision 
measurements of the pulses of such pulsars. Note that, at 
resonance, the neutron star Weber detector will exhibit {\it ringing} 
effect. Thus, even for a GW pulse, its 
effect will be present in the pulsar signal for a much longer duration allowing
folding of a large number of pulses to achieve desired accuracy. If
the source-pulsar distance is smaller, for example, when both are in the
same globular cluster, then the signal strength can be much larger.

The above estimates of changes of spin rate of NS have not
accounted for resonances for the increase of deformation amplitude. For this
discussion, we refer to the earlier paper \cite{pulsargw} only mentioning
the conclusion that there appears a good possibility that the neutron
star interior acts like a material of high quality factor $\bf{Q}$, and the
range of frequencies considered here are precisely in the range in which
typical neutron stars have various resonant modes. This should allow huge
enhancement in the amplitude of deformation, hence in the modification
in the pulses of the NS. For example,  for the Weber detector, 
in principle, one could get enormous amplitude enhancement at resonance 
with $Q$ factor of about $10^6$. Even for a short pulse, the ringing 
effect for Weber detector operating at resonance helps in 
tremendously enhancing signal to noise ratio. (Due to this ringing effect, 
the detector continues to vibrate in the resonant mode for some time even 
after the passing of the pulse through the detector. This time can be about
10 minutes for a GW burst lasting only a few ms, thereby allowing for 
separation between the random noise and the signal. In the
same way, if the pulsar continues to {\it ring} for some time after the 
GW pulse has passed through it, then the radio pulses will continue to
retain this {\it definite frequency} signal hidden within. Folding of many 
pulses may be able to separate this {\it ringing} signal.)

We had pointed out in \cite{pulsargw} that our proposal provides an 
opportunity to revisit the  gravitational wave events which have already 
been detected by LIGO/Virgo. For this one should look for specific 
pulsars which would have been disturbed by these events, and will transmit 
this disturbance via their pulse signals in any foreseeable future. If these
future pulsar events can be predicted with accuracy then a focused effort 
can be made to detect any possible changes in the signals of those specific
pulsars.  Any possible such detection will give valuable information about
the GW source as well as the neutron star parameters of the relevant pulsar
(such as equation of state). Interestingly, 
this also gives a possibility to actually detect those events whose GW signal 
detection has been missed. For example, many supernova events have been
identified within our galaxy from astronomical records around the world.
Any GW emission from these events (especially for type-II supernova
\cite{sngw} with estimated GW strain amplitude reaching as high as
$10^{-20}$ at a distance of 10 kpc from the source, see also, \cite{gwsn1a}) 
should have relatively strong effects on the galactic pulsars. Observing these
pulsars in future (or from past recorded pulse data) has the
possibility of identifying the signal of these supernova events
hidden in the pulsar signal. 

\begin{figure}
\begin{center}
 \includegraphics[width=0.8\linewidth]{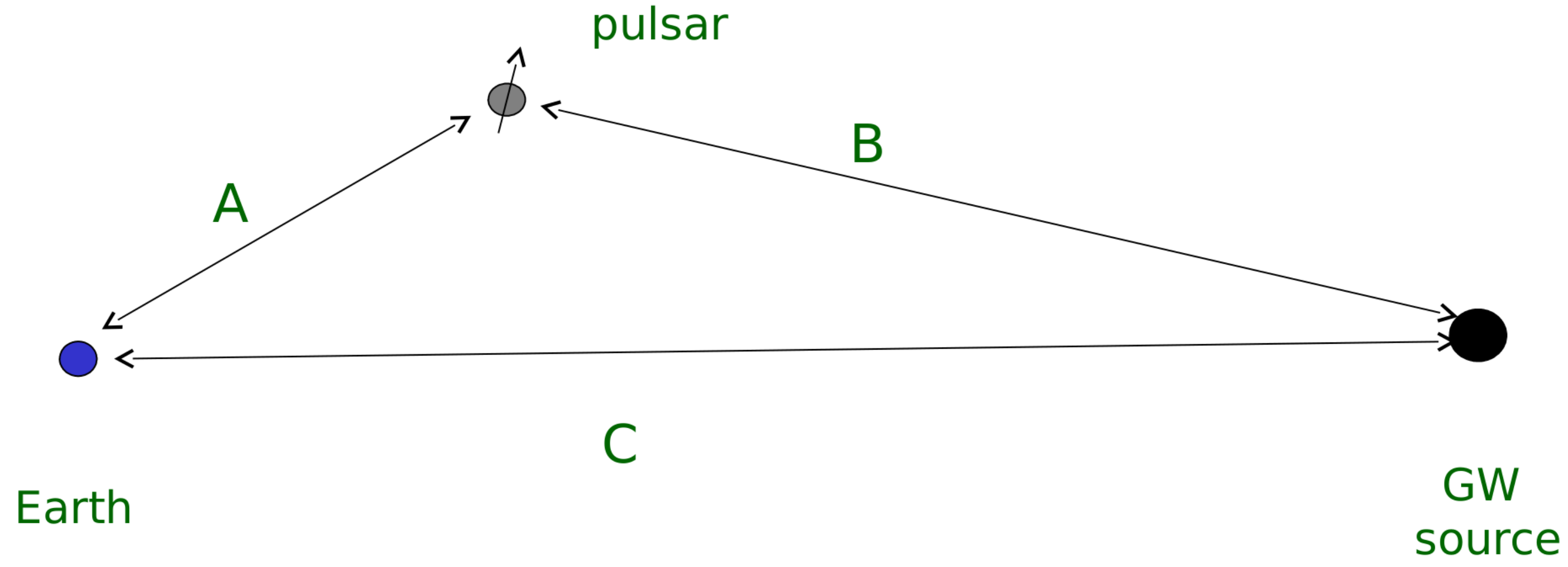}
 \caption{GW from the source reach earth directly via path C.
 These GW also reach a pulsar via path B affecting its structure, 
 hence its pulses. The affected pulses reach earth via path A.}
 \label{fig1}
 \end{center}
  \end{figure}

Fig.1 shows a typical GW event where the emitted GW from the source
reach earth directly via path C. These GW also reach a pulsar via 
path B affecting its structure, hence its pulses. The affected pulses 
reach earth via path A. The time when the affected pulses should reach
earth (measured from the time when GW event was directly seen/detected
on earth) is simply the light travel time for the path difference
$A + B - C$. We have taken the GW events from the catalogs \cite{gwtc,sndata}
and the pulsar data from the catalog \cite{pulsardata}, and have determined
the affected signal arrival date. To contain the size of the data
for presentation here we limit to the cases when the earliest signal 
arrival date is within next 50 years. We also include only those cases
where the error estimates (whenever available) limit the
uncertainty in the signal arrival date within additional 50 years.

\begin{table}[h]
	\scriptsize
    \centering
    \begin{tabular}{|c|c|c|c|c|}
        \hline
Source      &  Pulsar     &  Min. date & Mean &  Max. \\
            &             & mm/dd/year& year    &  year   \\
 \hline
           GW150914       & J0736-6304     &  4/24/2031 & 2031 & 2081 \\
\hline
            GW170814       & J0437-4715     &  1/14/2035 & 2038 & 2043 \\
\hline
            GW170817       & J1400-1431     &  5/21/2048 & 2048 & 2052 \\
\hline
            GW170818       & J2307+2225     & 12/ 2/2031 & 2037 & 2066 \\
\hline
           SN185          & J0900-3144     &  9/ 5/2033 & 2049 & 2066 \\
\hline
            SN1006         & B1944+17       &  4/ 7/2029 & 2032 & 2035 \\
                          & J0633+1746     &  4/10/2032 & 2032 & 2033 \\
                          & J0108-1431     & 11/28/2033 & 2034 & 2036 \\
                          & J2124-3358     & 10/23/2038 & 2044 & 2050 \\
\hline
            Crab          & J1856-3754     &  3/21/2057 & 2057 & 2058 \\
                          & J0919-42       &  6/21/2058 & 2076 & 2106 \\
\hline

 \multirow{2}{*}{SN1604} & J1734-3058     &  6/15/2019 & 2029 & 2040 \\
         & J1759-1956     &  3/ 1/2020 & 2039 & 2060 \\
         & J1736-3511     &  4/24/2025 & 2030 & 2035 \\
         & B1726-00       &  2/21/2027 & 2029 & 2031 \\
\cline{2-5}
	    & \multicolumn{2}{c}{J1738+04 (2027-2030),}{B1734-35 (2031-2040),} & & \\
	    & \multicolumn{2}{c}{J1800-0125 (2033-2037),}{J1832+0029 (2034-2036),} & & \\
	    & \multicolumn{2}{c}{J1911-1114 (2035-2038),}{B1800-21 (2035-2070),} & & \\
	    & \multicolumn{2}{c}{J1654-23 (2037-2064),}{J1851-0053 (2045-2048),} & & \\
	    & \multicolumn{2}{c}{J1609-1930 (2054-2060),}{B1732-02 (2062-2068),} & & \\
	    & \multicolumn{2}{c}{J1754-3510 (2063-2074),}{B1804-12 (2064-2078),} & & \\
	    & \multicolumn{2}{c}{J1758-2630 (2065-2101),}{J1717+03 (2067-2070).} & & \\
\hline
\multirow{2}{*}{ SN1885}  & J0139+33       & 11/12/2024 & 2024 & 2024 \\
                          & B2310+42       & 11/19/2030 & 2030 & 2030 \\
                          & J2302+4442     &  2/22/2031 & 2031 & 2031 \\
                          & J0103+54       &  4/ 1/2060 & 2060 & 2060 \\
                          & B2334+61       &  8/ 9/2064 & 2064 & 2064 \\
                          & J2307+2225     &  7/21/2065 & 2065 & 2065 \\
\hline
\multirow{2}{*}{SN1987A}  & J0709-5923     &  5/ 9/2023 & 2023 & 2023 \\
                          & B0559-57       &  7/ 8/2024 & 2024 & 2024 \\
                          & B0923-58       &  8/ 1/2024 & 2024 & 2024 \\
                          & J0621-55       & 11/15/2024 & 2024 & 2024 \\
\cline{2-5}
	    & \multicolumn{2}{c}{J0437-4715 (2028-2029),} {B0403-76 (2034),} &  & \\
	    & \multicolumn{2}{c}{J0656-5449 (2041),} {J1107-5907 (2050),}  & & \\
	    & \multicolumn{2}{c}{J1017-7156 (2050),} {B1055-52 (2052),}  & & \\
	    & \multicolumn{2}{c}{B1014-53 (2055),}{J0849-6322 (2060),}  & & \\
	    & \multicolumn{2}{c}{J1000-5149 (2062),}{B0901-63 (2064).}  & & \\
\hline

\end{tabular}
\caption{GW signal arrival dates for next 50 years. The GW sources and the
relevant pulsars are listed along with the earliest date expected for the 
perturbed pulsar signal to reach earth (Min. date), the mean value of the 
year (Mean year) and the maximum year (Max. year). The mean year value here 
corresponds to the specific values of the distance and the coordinates 
listed for the GW event and the relevant pulsar, while minimum and maximum 
dates are calculated using various errors given for these quantities 
(wherever available). We give the date of earliest signal arrival in days 
and months also even though the error estimates are in years. This is only 
meant to reflect the exactly known date of the source, and to illustrate the 
possibility that with improvements in various observations (source and 
pulsar locations), in principle, a very precise date of signal arrival on 
earth could be predicted.} 
\label{tab1}
\end{table}

TABLE I, presents the signal arrival dates for future in the next 50 years.
We list here the source of the gravitational wave and the pulsar whose
perturbed pulses will reach earth.  Keeping track of various errors wherever 
available (in the locations of the source and the pulsars) \cite{pulsarerr}, 
we give the earliest date expected for the signal to reach earth (Min. 
date), the mean value of the year (Mean year) and the maximum year 
(Max. year). The mean year value here corresponds to the specific values of
the distance and the coordinates listed for the GW event and the relevant 
pulsar, while minimum and maximum dates are calculated using the errors 
given for these quantities.  For few cases there was a long list of pulsars 
(even with the limit of 50 years for their signal arrival date on earth). (To 
save space, we have listed only few cases in detail and remaining entries
are listed in the format {\it pulsar (min year - max year)}, (e.g. for
SN1604, SN1885, and SN1987A). When max year and min year are the same then 
it is listed as {\it pulsar (year)}.) The error estimates are not
very exhaustive here, and more effort is needed to give more reliable
estimates for this. But the mean value is calculated with the mean
values for distances and coordinates available, giving clear idea
that many such events could be observed in very near future.
We give the date of earliest signal arrival in days and months also
even though the error estimates (from errors in source and pulsar
location) are in years. This is only meant to reflect the exactly known 
date of the source, and to illustrate the possibility that with improvements 
in various observations (source and pulsar locations), in principle, a very 
precise date of signal arrival on earth could be predicted. 
For example, for supernova SN1987A we find that perturbed signals 
from pulsars J0709-5923  and B0559-57 are expected to reach earth on 
5/9/2023 and 7/8/2024 respectively, with many more pulsar signals 
expected in subsequent years.  For Crab 
supernova, we find the signal arrival date from pulsar J1856-3754 in March 
2057 with error of one year (from the error in Crab source distance).
Among the merger events detected by LIGO/Virgo, GW170814 event will have
its imprints on the signals on the pulsar J0437-4715 reaching earth
between 2035-2043. Interestingly, even for the earliest observed supernova
SN185 (recorded in the year 185 AD), we have a possibility of observing
the original event imprinted on the signal of pulsar J0900-3144 which
will reach us sometime between 2033-2066.
 
In TABLE II, we have presented the data for those past events where
signal arrival date lies between 1967-2019. This is with the idea that
since the discovery of pulsars, whatever data may have been recorded,
in principle, it could be analyzed to identify and imprints of
those events. We have avoided listing those events for which the 
uncertainty (in terms of known errors in locations)
is larger than 50 years, even if the earlier dates are within next few
years. Including those cases makes the list too large for presentations.
Our purpose here is to show the feasibility of this proposal with few
test cases where signal arrival date is very likely within a couple
of decades.
\begin{table}[h]
        \scriptsize
    \centering
    \begin{tabular}{|c|c|c|c|c|}
        \hline
Source      &  Pulsar     &  Min. date & Mean &  Max. \\
            &             & mm/dd/year& year    &  year   \\
 \hline
SN185          & J2241-5236     &  2/ 1/1968 & 1985 & 2004 \\
             & J1858-2216     &  6/ 7/2015 & 2031 & 2049 \\
\hline
SN1006         & J0934-4154     &  3/21/1989 & 1997 & 2006 \\
               & J0621-55       & 11/27/1989 & 1994 & 1999 \\
               & B1118-79       &  5/ 3/1994 & 2011 & 2029 \\
               & J1729-2117     &  3/21/1999 & 2022 & 2048 \\
               & B1919+21       &  8/20/2006 & 2009 & 2012 \\
\hline
Crab           & J2322-2650     &  8/ 5/1996 & 2005 & 2017 \\
\hline
\multirow{2}{*}{SN1604}  & J1755-2534     & 12/ 8/1968 & 1983 & 1998 \\
               & J1758-1931     & 10/13/1970 & 1986 & 2004 \\
               & J1725-2852     & 11/23/1970 & 1983 & 1997 \\
               & J1756-2225     &  6/30/1971 & 1991 & 2013 \\
\cline{2-5}
        & \multicolumn{2}{c}{B1642-03 (1974-1977),} {J1744-3130 (1979-1993),}  & & \\
        & \multicolumn{2}{c}{J1105-43 (1984),}{B1919+21 (1984),}  & & \\
        & \multicolumn{2}{c}{B1944+17 (1992-1993),}{J1737-3102 (1994-2011),}  & & \\
        & \multicolumn{2}{c}{J1000-5149 (2004-2005),}{J1813-1246 (2004-2013),}  & & \\
        & \multicolumn{2}{c}{B1740-31 (2007-2023),}{J1759-3107 (2008-2021),}  & & \\
            & \multicolumn{2}{c}{B0538-75 (2009),} {B1612-29  (2010-2015),}  & & \\
            & \multicolumn{2}{c}{J1750-28 (2015-2041),}{J1741-3016 (2016-2038).}  & & \\
\hline

SN1885         & J0242+62       &  4/ 5/1969 & 1969 & 1969 \\
               & B0052+51       &  4/26/1987 & 1987 & 1987 \\
               & J0059+50       & 11/ 9/1995 & 1995 & 1995 \\
               & B0045+33       & 10/15/2000 & 2000 & 2000 \\
               & J0106+4855     &  8/13/2001 & 2001 & 2001 \\
\hline
\multirow{2}{*}{SN1987A} & J0540-7125     &  6/24/1988 & 1988 & 1988 \\
         & B0538-75       &  1/21/1990 & 1990 & 1990 \\
         & J0711-6830     &  4/ 1/1991 & 1991 & 1991 \\
         & J0749-68       & 12/26/1994 & 1994 & 1994 \\
\cline{2-5}
            & \multicolumn{2}{c}{J0736-6304 (1996),} {J0511-6508 (1996),}  & & \\
            & \multicolumn{2}{c}{J0834-60 (2007),} {J0457-6337 (2018).}  & & \\

\hline
\end{tabular}
    \caption{GW signal arrival dates for past 50 years.}
\label{tab2}
\end{table}

 From Fig.1 we also note that relative directions of the GW propagation and
 the pulsar spin will lead to different effects on the pulse timing and
 profile which can be analyzed to get information about source direction
 even with a single pulsar-detector observation. It also shows an 
 interesting possibility of directly observing any circularly polarized
 gravitational wave (which could arise in supernova explosion 
 depending on the pre-explosion hydrodynamics \cite{circpl})
 by its direct effect on the spin rate and the pulse profile of the pulsar.

 Main conclusion of our paper is that different GW events can be
observed again and again via pulsars which are affected by these GW events.
This gives a remarkable possibility of visiting past GW events which we 
missed, e.g. allowing us to observe the actual signal of past supernova events, 
especially in our galaxy. The most important message here is that the
dates which are predicted are completely model independent, simply following
from the coordinates of the source and the relevant pulsars. Model details
come in estimating the strength of the signal. We have argued that
the signal should be in the observational limit (especially when pulsar
deformations are in resonance). However, we know  that there
are many uncertainties about the interiors of neutron stars, 
one of the most important factor being the quality factor $Q$ affecting
the resonance effects. Irrespective of these concerns, we must acknowledge
that the dates when the perturbed signal reaches us can be determined
precisely (with proper error estimates), and on those
dates (or, during those periods) we must carry out detailed observations
of pulsar signals so that we do not miss out the great opportunity
of revisiting past events. If this proposal works, it will open up
an entirely new way of observing GW events in the sky.

\section{Acknowledgments}
We thank Sanatan Digal, Oindrila Ganguli, and Arpan Das
for very useful discussions.


\begin{thebibliography}{99}

\bibitem{gwtc} B. P. Abbott et al. (LIGO Scientific Collaboration 
and Virgo Collaboration), arXiv:1811.12907 

\bibitem{pulsargw}
A. Das, S. S. Dave, O. Ganguly, and A. M. Srivastava,
Phys.Lett. {\bf B791} (2019) 167.

\bibitem{nstarpt} P. Bagchi, A. Das, B. Layek, A. M. Srivastava, 
Phys.Lett. {\bf B747} (2015) 120.

\bibitem{sngw} C. D. Ott, Class. Quant. Grav. {\bf 26} (2009) 063001.

\bibitem{gwsn1a} D. Falta, R. Fisher and G. Khanna,
Phys. Rev. Lett. 106 (2011) 201103. 

\bibitem{sndata}  Weblink: www.messier.seds.org $>$ more $>$  mw\_sn. 

\bibitem{pulsardata} https://www.atnf.csiro.au/people/pulsar/psrcat/

\bibitem{SN87} R. M. West,A. Lauberts, H. E. Jorgensen, H. E. Schuster, 
Astronomy and Astrophysics, Vol. 177, p. L1-L3 (1987).

\bibitem{crab} D. L. Kaplan, S. Chatterjee, B. M. Gaensler, and J. Anderson,
	Astrophys. J. {\bf 677}, 1201 (2008)

\bibitem{pulsarerr} D. L. Kaplan et al, Astrophys. J. {\bf 789}, 119 (2014).
A. T. Deller, J. P. W. Verbiest, S. J. Tingay and M. Bailes, 
Astrophys.J. {\bf 685}, L67 (2008). B. P. Abbott et al. (LIGO Scientific 
Collaboration and Virgo Collaboration), arXiv:1811.12907.

\bibitem{hinderer2008} T. Hinderer, Astrophys.J. {\bf 677} (2008) 1216.

\bibitem{abbott2017} B. P. Abbott et al. (LIGO Scientific Collaboration 
and Virgo Collaboration), Phys. Rev. Lett. {\bf 119} (2017) 161101.

\bibitem{carrollbook} S. M. Carroll, Spacetime and Geometry: An 
Introduction to General Relativity, Addison Wesley, 2004.

\bibitem{circpl} K. Hayama, T. Kuroda, K. Kotake, and T. Takiwaki,
Mon.Not.Roy.Astron.Soc. {\bf 477} (2018) L96. 

\end{thebibliography}
\end{document}